\documentclass[aps,prl,preprint,groupedaddress]{revtex4-1}

\usepackage{graphicx}
\usepackage{dcolumn}
\usepackage{amsmath}
\usepackage{textcomp}
\usepackage{color}
\usepackage{braket}

\begin{document}


\title{Multi-electron dynamics in the tunnel ionization of correlated quantum systems}


\author{Maximilian Hollstein}
\author{Daniela Pfannkuche}
\affiliation{Universit\"at Hamburg, I. Institut f\"ur Theoretische Physik, Jungiusstra{\ss}e 9 , 20355 Hamburg, Germany}


\date{\today}

\begin{abstract}
The importance of multi-electron dynamics during the tunnel ionization of a correlated quantum system is investigated. By comparison of the solution of the time-dependent Schr\"odinger equation (TDSE) with the time-dependent configuration interaction singles approach (TDCIS), we demonstrate  the importance of a multi-electron description of the tunnel ionization process especially for weakly confined quantum systems. Within this context, we observe that adiabatic driving by an intense light field can even  enhance the correlations between still trapped electrons.
\end{abstract}


\maketitle

\section{Introduction} \label{sec:introduction}The ionization of closed shell atoms is impressively well-understood on the basis of single active electron approaches (SAE\citep{PhysRevA.38.778,Kulander1991523,PhysRevLett.62.524}) or effective one-particle theories as the time-dependent configuration interaction singles approach (TDCIS\cite{PhysRevA.74.043420,PhysRevLett.111.233005,TDCISDissipation}). Within these approaches, ionization is described by the ejection of a single
electron into the continuum while the residual electrons remain unaffected (i.e. they are only taken into account by a time-independent potential for the active electron (SAE) or they are kept residing in Hartree-Fock ground state orbitals of the field-free atom (TDCIS)).  However, in weakly confined quantum systems such as molecules or atom-like systems as semiconductor quantum dots in which the electron-electron interaction induces significant correlations between the trapped electrons \cite{doi:10.1021/jp953749i,PhysRevB.47.2244,Pfannkuche19936,Szafran1999185,Henderson2001138}, an independent particle description of the ionization process, as inherent in these approaches, is expectably unsuitable from the very beginning. In this paper, we address the question concerning the importance of a multi-electron description of the tunnel ionization process of weakly confined and correlated quantum systems. For this purpose, we consider the dynamics which is induced by an intense low-frequency light field in a two-electron model system. Our conclusions drawn from these considerations, however, are not only valid for two-electron quantum systems but they are transferable to systems with more than two weakly bound electrons. Furthermore, we would like to remark that the conclusions drawn from the presented model calculations are not limited to the ionization of molecular or atomic systems but they are also relevant for experiments based on the application of a (time-dependent) voltage to semiconductor quantum dots \cite{PhysRevLett.72.1076,PhysRevB.81.165318} and photo assisted tunneling (PAT) \cite{PhysRevB.37.4201,PhysRevLett.72.1076,PhysRevLett.78.1536,PhysRevLett.109.077401,photoassistedlongrangetunneling} in the low-frequency regime. 
\newline\newline This paper is structured as follows: after introducing the considered model system, we demonstrate the need for a multi-electron description of the tunnel ionization process by comparison of the solution of the time-dependent Schr\"odinger equation (TDSE) with the results obtained by the TDCIS approach. By detailed analysis of the exact wavefunction we provide insight into the light induced dynamics revealing that a multi-electron description becomes necessary not only due to ground state correlations but also due to a light induced collective electron motion which is accompanied with an enhancement of 
correlations between still trapped electrons. \section{Model}\label{model} In order to study the ionization of a weakly confined quantum system,  we consider a one-dimensional model system consisting of two electrons in an inverse Gaussian confining potential. (Note that the effective potential realized in semiconductor quantum dots can be well approximated by this potential \cite{9783540636533,Gauss_potential_PhysRevB.62.4234}).
 Without exception, atomic units are used throughout this paper. Referring to the spatial coordinate as x, the considered confining potential can be denoted by:
\begin{equation}
\label{eq:confinement}
V(x) = -V_0 e^{-(\frac{x}{w})^2}
\end{equation} where $V_0$ determines the depth and $w$ the width of the confinement. Within the harmonic approximation of this potential, the strength of the confinement i.e. the level spacing of the lowest bound states is given by:\begin{equation}
\Delta E\approx \frac{\sqrt{2V_0}}{w}
\end{equation} In the subsequent considerations we make use of this dependency and vary the strength of the confinement by variation of the width $w$.
  \newline
The electron-electron interaction is taken into account by a regularized Coulomb interaction:
\begin{equation}
V_{ee}(x_1,x_2) =\frac{1}{ \sqrt{(x_1 - x_2)^2 + \delta^2}} 
\end{equation} where $x_{1,2}$ denote the spatial coordinates of the two electrons.  By choosing $\delta$ unequal zero (we set $\delta$ to 0.5), a finite width of the electronic wave function along the unconsidered spatial dimensions is taken into account in a phenomenological way\cite{PhysRevB.47.16353,0953-8984-18-9-002}.
For the results presented below, the depth of the confining potential  is set to $V_0 = 3$.\newline\newline Hereinafter, we restrict our considerations to light fields with a wavelength which is large compared to the extend of the considered quantum system. This allows the application of  the dipole approximation to the light-matter interaction.  Thus, in length gauge, the Hamiltonian of the quantum system interacting with the light field is given by:
\begin{equation}
\hat{H}_0 = -\frac{1}{2}\sum_{i = 1}^{2}\frac{\partial^2}{\partial x_i^2} -V_0\sum_{i = 1}^{2}  e^{-(\frac{x_i}{w})^2} + \frac{1}{ \sqrt{(x_1 - x_2)^2 + \delta^2}} 
\end{equation}
\begin{equation}
\hat{H}(t) = \hat{H}_0 - \sum_{i = 1}^{2}x_iE(t)
\end{equation}
where $\hat{H}_0$ is the Hamiltonian of the field free system and E(t) denotes the time-dependent electric field of the laser. Subsequently, we consider the initial dynamics of the system in an electric field with sinusoidal time dependence i.e.: \begin{equation}E(t) = E_0 \sin(\omega t).\end{equation} \newline
In the following, we consider the situation where the two-electron system is initially prepared in the singlet ground state. In order to study the light induced dynamics, we solve the time-dependent Schr\"odinger equation numerically. For this, we employ a split operator technique \cite{Feit1982412}. Reflections at the boundaries of the  finite simulation box are avoided by the use of a complex absorbing potential $V_{CAP}$.
\begin{equation}
\label{eq:CAP}
V_{CAP} =
			 \begin{cases}
     -iC(||x|-x_0|)^3 & |x| >x_0 \\
     0 & else
   \end{cases}
\end{equation}
For the calculations presented below, C is chosen to $4\times 10^{-3}$. $x_0$ is chosen to 15 where both the confining potential and the ground state wavefunction are only insignificant different from 0 (i.e. for all model parameters, the absolute value of confining potential is for x = 15 smaller than 0.001 and the value of the ground state density is decreased to $\lesssim 10^{-16}$ ).\newline In this paper, we consider the tunnel ionization process i.e. the ionization of the considered quantum system via tunneling through the effective potential barrier which is formed by the instantaneous electric field and the confining potential. Considering that the tunnel ionization  regime is  characterized by a Keldysh parameter \cite{Keldysh} much smaller than one, we choose the frequency of the light field to $\omega = 2\pi \times 0.001$ and the amplitude to $E_0=0.3$ corresponding to a Keldysh parameter smaller than 0.01 for the considered model potential parameters. For semiconductor quantum dots, the characteristic confinement energy is in the meV range so that the considered light field would be in the domain of far infrared light to microwaves. We found that then, excitations within the potential well are negligible for the considered laser parameters so that we here consider subsequently only the quasi-adiabatic dynamics during a half-cycle of the laser field. Since for tunnel ionization, the ionization process is most prominent at times when the electric field of the light field is extremal, we focus especially on the situation at t = T/4 when the electric field is maximum. Here, we restrict our considerations to the dynamics in the confining potential well and its vicinity  by analyzing the restricted wavefunction: \begin{equation}\tilde{\Psi} = \frac{\Psi(x_1,x_2,t)|_{|x_{1,2}|< x_{max}}}{\sqrt{ \int_{ -x_{max}}^{x_{max}}\int_{ -x_{max}}^{x_{max}} dx_1dx_2|\Psi(x_1,x_2)|^2} }\end{equation} whereas unless noted differently, we chose 
\begin{equation} 
	x_{max} = x_0 = 15.
\end{equation} Before the two-electron system is exposed to the light field, both electrons are situated in this area. However, at later times, electron(s) are excited by the laser into the continuum  and leave the considered spatial region. We regard these electrons as ionized and determine the ionization probability consequently by:\begin{equation}
P_{ion}(t) = 1 - \int_{ -x_{max}}^{x_{max}}\int_{ -x_{max}}^{x_{max}} dx_1dx_2|\Psi(x_1,x_2,t)|^2.\label{eq:pion}\end{equation} \newline 
\section{Results}\subsection{Comparison of TDCIS and TDSE}
\begin{figure}[h!]
  \centering
\includegraphics[scale=0.45]{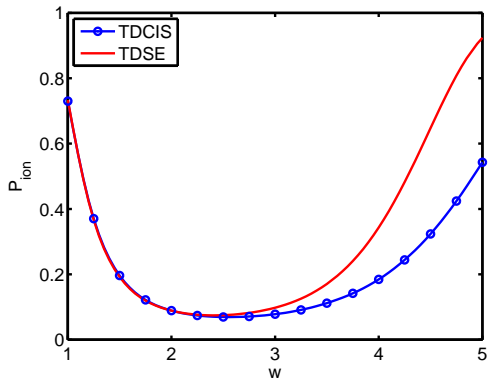}
\caption{(Color online)
The ionization probability $P_{ion}$  as defined in equation \ref{eq:pion} after a half cycle of the laser field obtained by the solution of the TDSE (red curve) and by the TDCIS approach (blue curve with dots) in dependence on the width of the confinement.  For narrow confinements (i.e. for $w < 2.5$), the TDCIS approach reproduces very accurately the exact ionization probability whereas for wide confinements substantial deviations are observable.
}\label{ionprobability}
\end{figure}

\begin{figure}[h!]
 \centering
\includegraphics[scale=0.45]{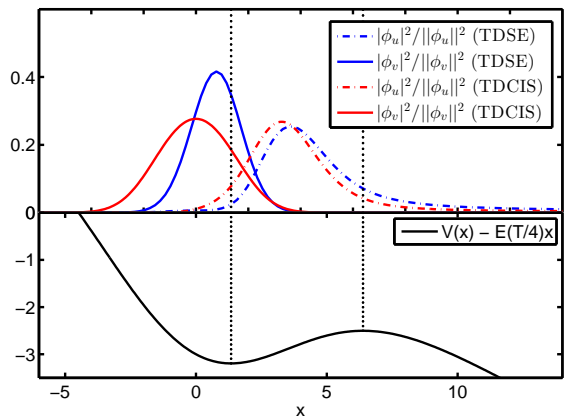}
 \caption{(Color online)
The renormalized orbitals which constitute the (approximate) open-shell form of the exact wavefunction and the TDCIS wavefunction at $t = T/4$ for a confining potential width $w =  5$. Both the exact and the TDCIS wavefunction are described by one orbital which is localized in the potential well $\phi_v$ and respectively $\psi_v$ (solid curves) and one orbital which has a significant overlap with the tunneling barrier describing an tunneling electron $\phi_u$ and respectively $\psi_u$ (dashed curves). Both orbitals obtained from the exact wavefunction are shifted to the right  with respect to the corresponding orbitals obtained from the TDCIS wavefunction.\label{open_shell}
}
\end{figure}

Within the TDCIS approach, the two-electron wavefunction is expanded  in the Hartree-Fock ground state $\ket{\Phi_0}$
and its particle-hole excitations $\ket{\Phi^i_0} $
\begin{equation}
\ket{\Psi} =\alpha_0\ket{\Phi_0} + \sum_{i > 0}\alpha_0^i\ket{\Phi^i_0} 
\end{equation}
with
\begin{equation}
\ket{\Phi^i_0} =\frac{1}{\sqrt{2}} (c^{\dag}_{i\uparrow} c_{0\uparrow} + c^{\dag}_{i\downarrow} c_{0	\downarrow})\ket{\Phi_0}
\end{equation}

where $c_{0\uparrow}, c_{0\downarrow}$ denote annihilation operators of the spin orbitals which are occupied in the Hartree-Fock ground state determinant while $c^{\dag}_{i\uparrow}, c^{\dag}_{i\downarrow}$  denote creation operators of virtual orbitals. The spatial part of the TDCIS singlet two-electron wavefunction is consequently given by:
\begin{equation}
\label{eq:TDCISspatial}
\Psi(x_1,x_2,t)_{_{TDCIS}} = \sum_{i} \tilde{\alpha}_i (t)\big[\psi_0(x_1) \psi_i(x_2) + \psi_i(x_1) \psi_0(x_2)\big]
\end{equation} 
with
\begin{equation}
\tilde{\alpha}_i (t) = 			 \begin{cases}
     \alpha_0(t)/2 & i = 0\\
     \alpha_0^i(t) & else.
   
   \end{cases}
\end{equation}
As can be seen in equation (\ref{eq:TDCISspatial}),  at all times, only one electron can be active (i.e. occupy an arbitrary orbital) while the other electron is forced to occupy the Hartree-Fock ground state orbital $\psi_0$.\\In order to determine the coefficients $\tilde{\alpha}_i(t)$ we solve the Hartee-Fock equations on a pseudo-spectral grid as described in \cite{Greenman2010}. The TDCIS wavefunction is then propagated by iterative Lanczos reduction \cite{Park1986} within the configuration interaction singles singlet subspace constructed by the eigenfunctions of the Fock-operator.\newline In figure \ref{ionprobability}, the ionization probability as defined in eq. \ref{eq:pion} obtained by the exact solution of the TDSE and by the TDCIS approach are shown for varying confinement widths. As can be seen, the TDCIS approach reproduces very accurately the exact ionization probability for  $w < 2.5$ whereas large deviations are observable for $w > 2.5$ (e.g. by 40 percent for $w = 5$).  \newline Some insight can be gained by the approximation of the exact two-electron wavefunction in terms of configurations constructed from the two most occupied natural orbitals (cf. eq. \ref{formula_truncated_naoCI}). This truncated configuration interaction expansion allows a good approximation of the exact wavefunction since the naturals orbitals constitute an orbital basis set leading to the most rapidly converging expansion in configurations \cite{Coulomb_energy_dot_width_PhysRevB.43.7320} 
\begin{equation}\Psi \approx \sum_{i,j = 1}^2c_{ij}\phi_i(x_1)\phi_j(x_2)/\sqrt{\sum_{i,j = 1}^2|c_{ij}|^2}\label{formula_truncated_naoCI}.\end{equation}
If the coefficients $c_{ij}$ are real, this approximate wavefunction  can be represented exactly in open shell form \citep{NaturalOrbitalsintheQuantumTheoryofTwoElectronSystems} where one electron occupies a spatial orbital $\phi_u$ while the other electron occupies a different orbital $\phi_v$, which is not nessecary completely orthogonal or totally parallel to $\phi_u$:
\begin{equation}\Psi(x_1,x_2,t) \approx \phi_u(x_1,t)\phi_v(x_2,t) + \phi_v(x_1,t) \phi_u(x_2,t).\label{eq_open_shell_form}\end{equation} 
Although the coefficients $c_{ij}$ of the approximate wavefunction (\ref{formula_truncated_naoCI}) obtained from the numerical propagation are complex at $t = T/4$, we found for this situation that with good accuracy $c_{00}$ and $c_{11}$ can be chosen real with alternating sign. Since $c_{01}$ and c$_{10} $ vanish exactly (\cite{Giesbertzphd,PhysRevA.88.052514}), a representation in open shell form is nonetheless possible. As described in reference \citep{NaturalOrbitalsintheQuantumTheoryofTwoElectronSystems}, $\phi_u$ and $\phi_v$ can be chosen as: 
\begin{equation}
\phi_u =(|c_{00}|^{\frac{1}{2}}\phi_0 +| c_{11} |^{\frac{1}{2}}\phi_1) / (4(|c_{00}|^2 + |c_{11}|^2) )^{\frac{1}{4}}
\end{equation}
\begin{equation}
\phi_v =(|c_{00}|^{\frac{1}{2}}\phi_0 - |c_{11} |^{\frac{1}{2}}\phi_1) / (4(|c_{00}|^2 + |c_{11}|^2) )^{\frac{1}{4}}.
\end{equation}
With this, both  wavefunctions (eq. \ref{formula_truncated_naoCI} and eq. \ref{eq_open_shell_form}) have an overlap of 0.967 with the exact wavefunction and thus provide a reasonable approximation. Now, the approximation of the exact wavefunction in open-shell form allows a convenient comparison with the TDCIS wavefunction which can be exactly represented in open-shell form:\begin{equation}
\Psi(x_1,x_2,t)_{_{TDCIS}} = \psi_u(x_1,t)\psi_v(x_2,t) + \psi_v(x_1,t)\psi_u(x_2,t)
\end{equation}
where:
\begin{equation}
\label{psiuequalspsi0}
\psi_v(t) = \psi_0
\end{equation}
and
\begin{equation}
\psi_u(t) = \sum_{i} \tilde{\alpha}_i (t)\psi_i.
\end{equation}
The two orbitals $\phi_u$ and $\phi_v$ and respectively $\psi_u$ and $\psi_v$ are only fixed up to a factor i.e. a two-electron wavefunction in open shell form is invariant under the transformation $\phi_u \rightarrow \phi_u/\alpha$, $\phi_v \rightarrow \alpha\phi_v $ where $\alpha$ is an arbitrary non-zero complex number. Therefore, we compare the TDCIS wavefunction with the exact wavefunction by consideration of the corresponding renormalized orbitals (cf. figure \ref{open_shell}). As can be observed in figure \ref{open_shell} for $w = 5$, within this open shell approximation, both the exact and the TDCIS wavefunction are described by one orbital which is localized in the potential well ($\phi_v$(TDSE) and respectively $\psi_v$(TDCIS)) and one orbital which has a significant overlap with the tunneling barrier describing an tunneling electron ($\phi_u$(TDSE) and respectively $\psi_u$ (TDCIS)). This indicates that the dominant ionization process is also within the exact treatment a process where one electron resides in the potential well while the other is tunneling through the potential barrier. For TDCIS, the localized orbital (i.e. $\psi_v$) coincides with the Hartree-Fock ground state orbital $ \psi_0$ (cf. eq. \ref{psiuequalspsi0}) which is centered at the origin ($\braket{x}_{\psi_v} = \frac{\int|\psi_v|^2dx}{||\psi_v||^2} = 0$) while the corresponding orbital of the open shell approximation of the exact wavefunction $\phi_v$ ($\braket{x}_{\phi_v} = 0.862$) is shifted towards the local minimum of the instantaneous potential  at $x = 1.344$. Significant deviations are also observable between the orbitals describing the tunneling electron i.e. $\psi_u$  for the TDCIS approach and $\phi_u$ for the open shell approximation of the exact wavefunction.  In particular the orbital $\phi_u$ (TDSE) ($\braket{x}_{\phi_u} =4.829$) is shifted towards the potential barrier maximum at $x = 6.385$ with respect to $\psi_u$ (TDCIS) ($\braket{x}_{\psi_u} =4.2026$) resulting in a larger overlap of $\phi_u$ (TDSE) with the potential barrier.
This indicates that the immobility of the localized electron within the TDCIS approach effects via the Coulomb interaction also the tunneling electron. That is, within the TDCIS approach, the tunneling electron is less effectively pushed towards the tunneling barrier. This provides an explanation for the  considerable differences between the ionization probability obtained by the TDCIS approach and the exact treatment. That is, in particular the underestimation of the ionization process by the TDCIS approach can be related to this circumstance(cf. figure \ref{ionprobability}). \newline So far only an open shell approximation of the exact wavefunction is considered which approximates the full wavefunction fairly well (i.e. the overlap with the exact wavefunction is 0.967) but obviously there are still some deviations to the exact wavefunction. In the following section we therefore provide more extensive insight into the light induced multi-electron dynamics from a different perspective by considering the complete wavefunction obtained from the exact treatment.
\subsection{Field-Induced Multi-Electron Motion}
\begin{figure}[h!]
  \centering
\includegraphics[scale=0.45]{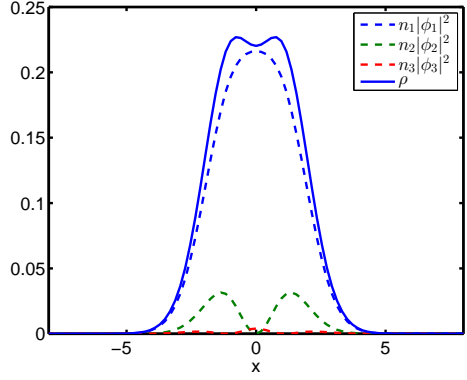}
\caption{(Color online)
The probability density $\rho$ of the singlet ground state and the densities of the first three most occupied natural orbitals $\phi_1,\phi_2$ and $\phi_3$ weighted by their occupation numbers ($n_1$,$n_2$ and $n_3$) for a relatively wide confining potential (w = 5). Due to a strong Coulomb interaction with respect to the confinement energy,  more than one natural orbital is significantly occupied in the ground state.
}\label{densitynaost0}
\end{figure}

\begin{figure}[h!]
\centering
\includegraphics[scale=0.45]{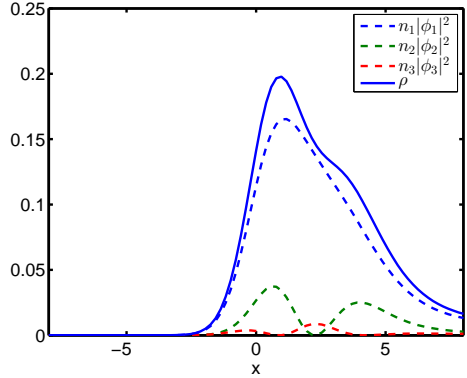}
\caption{(Color online) The probability density $\rho$ of the two electron state at the time when the electric field of the laser is strongest (t = T/4)  and the densities of the first three most occupied natural orbitals $\phi_1,\phi_2$ and $\phi_3$ weighted by their occupation numbers for a relative wide confining potential ($w = 5$). In comparison to the ground state (cf. figure \ref{densitynaost0}), the occupation of the second and the third most occupied natural orbital is increased.
}
\label{densitynaostT4}
\end{figure}

\begin{figure}[h!]
  \centering
\includegraphics[scale=0.45]{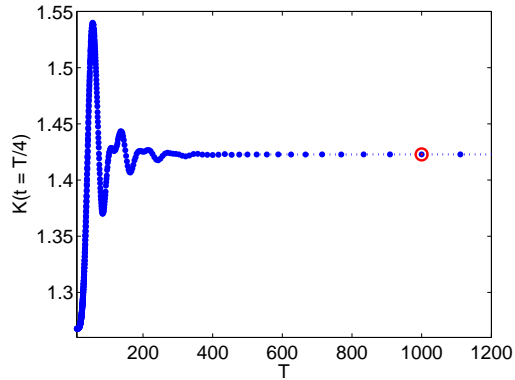}
\caption{(Color online) The degree of correlations K (cf. eq. \ref{eq:Kcorrelations}) at $t = T/4$ for $w = 5$. Non-adiabatic excitations within the potential well manifest themselves by a frequency dependence of K for $T < 400$ i.e. for frequencies larger than $2\pi\times0.0025$. Thus, for the considered laser frequency ($2\pi\times0.001$ - cf. red marker) non-adiabatic effects appear to be irrelevant.
}
\label{n3_vs_frequency}
\end{figure}

The low-frequency field considered here induces a a quasi-adiabatic electron motion leading to a time-dependent shift of the electronic center-of-mass. Within the harmonic approximation of the confining potential and within the dipole approximation for the light field, this shift is given by:
\begin{equation} \Delta x = \frac{E(t)w^2}{2V_0}.
 \end{equation} 
Hence, the amplitude of the collective two-electron motion increases monotonically with increasing width of the potential well (i.e. $\propto w^2$). Noteworthy, this shift of the potential minimum results in the motion of both electrons which becomes noticeable in the open shell approximation (cf. eq. \ref{eq_open_shell_form}) by the fact that the orbital $\phi_v$, which describes a localized electron, is not centered at the origin but is rather shifted towards the local minimum of the instantaneous potential (see fig. \ref{open_shell}). The regime of large deviations of the ionization probability obtained by TDCIS and exact treatment coincides with the regime where $\Delta x$  is in the order of the Bohr radius or larger i.e. $\Delta x \gtrsim 1     $ $\widehat{=}$  $w \gtrsim 4.5$. This supports the explanation for the large deviations between the ionization probability obtained by TDCIS and exact treatment as given in the previous section.  That is, within the TDCIS approach, the tunneling electron is less effectively pushed towards the tunneling barrier due to the immobility of the localized electron. \newline However, since the considered confining potential well is not a pure harmonic potential also excitations of the relative motion are possible \cite{9783540636533}. This allows for correlated electron dynamics. 
In order to get insight into this light-induced correlation dynamics, we consider electronic correlations which become noticeable by the circumstance that the electronic wavefunction can not be represented by a single Slater determinant. Therefore, we determine the degree of correlation of the two-electron wavefunction by employing the measure of correlations $K$ as defined in \cite{Eberly_PhysRevA.50.378}:
\begin{equation}
 \label{eq:Kcorrelations}
K =(\sum_i{n_i^2})^{-1}
\end{equation}
where $n_i$ denote the occupation numbers of the natural orbitals i.e. the eigenvalues of the first order density matrix \cite{QuantumTheoryOfCohesivePropertiesOfSolids}.
Thereby, $K$ can be interpreted as the "number" of Slater determinants which are effectively necessary to represent the wavefunction (cf. \cite{Eberly_PhysRevA.50.378}). Since every fully uncorrelated singlet two-electron wavefunction is a Slater determinant with a doubly occupied spatial orbital, the absence of correlations is characterized by a measure of correlations K equals one. Thus, electronic correlations manifest themselves by a value of K larger than one (Note that $n_i > 0$ and $\sum_i{n_i} = 1$ cf. \cite{QuantumTheoryOfCohesivePropertiesOfSolids} and \cite{NaturalOrbitalsintheQuantumTheoryofTwoElectronSystems}). In order to determine K numerically, we obtain the occupation numbers of the natural orbitals by diagonalization of the first order density matrix \cite{NaturalOrbitalsintheQuantumTheoryofTwoElectronSystems} represented on a spatial grid. \begin{figure}[h!]
 \centering
\includegraphics[scale=0.45]{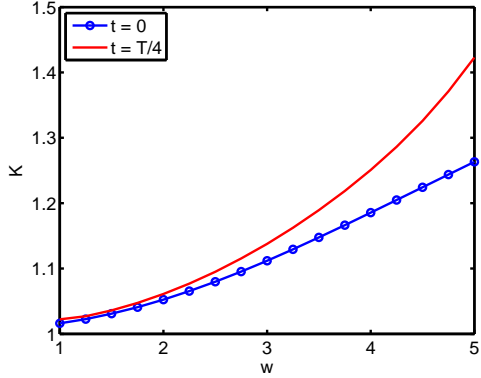}
 \caption{(Color online) The degree of correlations K as defined in eq. \ref{eq:Kcorrelations} for the singlet ground state (blue curve with dots) and at the time when the electric field of the laser is strongest (t = T/4) (red curve). Since the relative strength of the Coulomb interaction increases with increasing width of the confining potential, the amount of correlations present in the ground state increases correspondingly (cf. blue curve with dots). While for narrow confinements ($w \approx 1$), the amount of correlations remains constant, for broad confinement potentials, in presence of the light field, the amount of correlations is significantly increased in comparison to the ground state.
} \label{correlations}
\end{figure}\newline In figure \ref{correlations}, K is shown in dependence on the width $w$ of the confining potential for two situations i.e. for the singlet ground state (blue curve) and for the two-electron wavefunction at $t = T/4$ (red curve). As can be seen in figure \ref{correlations}, for narrow confinements ($w \approx 1$) the measure of correlations K for the singlet ground state  is nearly one indicating an accurate description by the Hartree-Fock determinant (for $w = 1$ one finds indeed that mostly only one natural orbital is populated in the ground state with an occupation of $0.992$ ). However, K increases monotonically with the confining potential width. This is related to a feature well known for the harmonic confinement potential, namely, that the confinement energy and the Coulomb interaction scale differently with respect to the characteristic confinement length $l_0$. That is, whereas the confinement energy scales as  $\frac{1}{l_0^2}$, the Coulomb interaction is proportional to  $\frac{1}{l_0}$. Hence, the relative strength of the Coulomb interaction increases monotonically with the width of the confinement potential leading to a correspondingly increasing population of more than one natural orbital (cf. figure \ref{densitynaost0}). For $w = 5$, one finds that a second natural orbital is significantly populated with an occupation of $0.107$ so that here, a multi-determinant treatment is already necessary to represent the ground state accurately (Note that the overlap with the Slater determinant with maximum overlap with the exact wavefunction is given by $\sqrt{n_1} = 0.9398$ where $n_1$ is the occupation number of the most occupied natural orbital (cf. \cite{NaturalOrbitalsintheQuantumTheoryofTwoElectronSystems}) ).  However, if one considers the two-electron wavefunction at $t = T/4$, one observes a significant enhancement of correlations with respect to the ground state correlations. For instance for $w = 5$, the occupation of the second most occupied natural orbital increases from $0.107$ to even $0.142$ in the presence of the light field and furthermore also a third natural orbital becomes noticeably populated with an occupation of $\approx 0.028$
 whereas its ground state occupation is only $0.009$ (cf. figure \ref{densitynaostT4}). Note that the occupation of a third natural orbital is not included in the open shell approximation (cf. eq. \ref{eq_open_shell_form}) and thus represents an additional deviation between TDCIS wavefunction and exact wavefunction which comes along  those discussed in the previous section. 
 Since the degree of correlations at $t = T/4$ is independent of the laser frequency for frequencies smaller than $2\pi \times 0.0025$ (cf. fig.  \ref{n3_vs_frequency}), non-adiabatic excitations within the potential well are appearently not the reason for the enhancement of  K (the considered laser frequency is $2\pi \times 0.001$). Since the most occupied natural orbitals are localized within the well of the instantaneous potential, this effect can be attributed to a quasi-adiabatic rearrangement of the still trapped electrons within this strongly deformed potential well. The enhancement of the degree of correlations indicates here an effective broadening of the potential well by the light field which results in an enhanced effective strength of the Coulomb interaction between the still trapped electrons. \newpage
\section{Conclusions} \label{sec:conclusions}
To conclude, we demonstrated here that for an accurate description of the tunnel ionization process, a multi-electron description is the more important the weaker the confining potential. By analysis of the exact wavefunction, we show that a multi-electron description is not only necessary due to ground state correlations but also due to a collective and correlated multi-electron motion  resulting from a strong deformation of the confining potential well by the light field.

\bibliography{references}

\end{document}